\documentclass[useAMS,usenatbib]{mn2e}
\usepackage{graphicx,amssymb,color}
\usepackage[normalem]{ulem}
\newcommand{\rev}{ }

\title[Post-MS fate of HR 8799]
{The post-main-sequence fate of the HR 8799 planetary system}

\author[]{Dimitri Veras$^{1,2}$\thanks{E-mail: d.veras@warwick.ac.uk}\thanks{STFC Ernest Rutherford Fellow},
Sasha Hinkley$^{3}$
\\
$^{1}$Centre for Exoplanets and Habitability, University of Warwick, Coventry CV4 7AL, UK
\\
$^{2}$Department of Physics, University of Warwick, Coventry CV4 7AL, UK
\\
$^{3}$Astrophysics Group, University of Exeter, Physics Building, Stocker Road, Exeter EX4 4QL, UK
}

\pubyear{2021}

\begin{document}
\label{firstpage}
\pagerange{\pageref{firstpage}--\pageref{lastpage}}
\maketitle

\begin{abstract}
The noteworthy four-planet HR 8799 system teeters on the brink of gravitational instability and contains an A-type host star which is characteristic of the progenitors of the majority of known white dwarf planetary system hosts. \cite{gozmig2020} have demonstrated that the system can retain all four planets for at least 1 Gyr along the main sequence if the planets evolve within an externally unperturbed $8$:$4$:$2$:$1$ mean motion resonance configuration. Here we propagate forward their most stable fit beyond the main sequence, and incorporate external effects from Galactic tides and stellar flybys. We find that (i) giant branch mass loss always breaks the resonance, and usually triggers the ejection of two of the planets, (ii) stellar flybys and Galactic tides rarely break the resonance during the main-sequence and giant branch phases, but play a crucial role in determining the final planetary configurations around the eventual white dwarf host star, and (iii) the meanderings of the surviving planets vary significantly, occupying regions from under 1 au to thousands of au. The ubiquitous survival of at least one planet and the presence of the debris discs in the system should allow for dynamical pathways for the white dwarf to be metal-polluted.  
\end{abstract}


\section{Introduction}

Over a decade after its discovery \citep{maretal2008,maretal2010}, the HR 8799 planetary system remains a benchmark of exoplanetary science. The multifaceted appeal of the system -- the first imaged multi-planet exosystem, and one that might be in the process of destabilizing -- arises from both its observational and dynamical properties.  

\subsection{Observations of HR 8799}

\subsubsection{Planets}

The HR\,8799 system is comprised of a $\lambda$\,Boo A5V host star with close proximity to the Sun \citep[41.29$\pm$0.15\,pc,][]{gaietal2018}. Three co-moving companions were initially identified by \citet{maretal2008} at projected orbital separations of 68, 38 and 24\,au over multiple epochs using adaptive optics and coronagraphy at the Gemini and W.M.~Keck Observatories.  A host of age indicators for HR\,8799, including its galactic space motion (indicating a high likelihood of being a member of the Columba moving group) and the placement of the star in a colour-magnitude diagram suggest a best-estimate age of $\approx$30\,Myr \citep{zucetal2011,maletal2013}. 

Based on this age estimate, as well as the measured near-infrared brightnesses of the co-moving companions, \citet{maretal2008} assigned mass estimates of 7$_{-2}^{+4}$, 10$_{-3}^{+3}$, and 10$_{-3}^{+3}$\,$M_{\rm Jup}$ for the HR\,8799bcd planets, respectively. Subsequent observations of this system at the more favourable observing wavelength of 3.8\,$\mu$m \citep{maretal2010} unveiled the direct detection of a fourth planet (``e'') at a significantly smaller projected orbital separation of 14.5$\pm$0.4\,au, and an estimated mass of 7$_{-2}^{+3}$\,$M_{\rm Jup}$, similar to that of the ``c'' and ``d'' companions.   

The relatively wide angular separations on the sky (ranging from $\approx 0.4 - 1.7^{\prime\prime}$) and high intrinsic luminosity of the four planets has allowed for some of the most detailed characterization of the planets' atmospheres through \textit{direct} spectroscopy using instruments that are dedicated to exoplanet imaging \citep[e.g.~GPI, SPHERE,][]{ingetal2014,zuretal2016,greetal2018}. Also, multi-purpose instruments equipped with much higher spectral resolution \citep[$\lambda/\Delta\lambda$$\sim$4000,][]{konetal2013} have indicated the presence of vertical atmospheric mixing when combined with detailed atmospheric models \citep{baretal2011}. More recently, the GRAVITY instrument at the Very Large Telescope has revealed the ability of using interferometric measurements to measure a spectrum with resolution $R\sim500$, a value which is higher than that of typical dedicated exoplanet imaging instruments such as GPI and SPHERE.  The GRAVITY measurements returned precise atmospheric composition measurements of HR\,8799~e, allowing for the derivation of a remarkably precise value of the atmospheric Carbon-to-Oxygen ratio of 0.60$\pm$0.08 \citep{moletal2020}, suggesting that the planet formed outside of the CO$_2$ or CO ice line. 

The 13-year time baseline since the discovery of the HR\,8799 planets has allowed for careful astrometric monitoring of the planets' orbits with instruments that achieve high contrast \citep[e.g.][]{maietal2015,pueetal2015,konetal2016,wanetal2018}.  Nevertheless, some of the tightest astrometric constraints originate from 1998 \textit{Hubble Space Telescope} observations of the system.  Remarkably, when novel image processing techniques were applied to this 1998 dataset \citep{souetal2012}, the locations of the three outermost planets HR\,8799~bcd \citep{souetal2011} were revealed, providing a $\gtrsim$20 year time baseline for astrometric monitoring. Most recently, GRAVITY has returned astrometry of the ``e'' planet relative to the star with precision of $\approx$100$\mu$as, placing strong constraints on the orbit, and disfavouring perfectly coplanar orbits \citep{graetal2019}.

\subsubsection{Debris belts}

In addition to containing four planets, the system also hosts belts of debris. Following the early identification with \textit{IRAS} of HR\,8799 being dust-rich \citep{sadnis1986}, more recent observations of this system in the mid-infrared \citep{matetal2014} and submillimetre \citep{holetal2017} have provided a more complete picture of the underlying structure of the population of dust and planetesimals.  

Specifically, by using spectroscopy from the \textit{Spitzer Space Telescope}, \cite{suetal2009} were able to identify a debris structure comprised of two components. The first is a warm dust component ($T\approx 150$\,K) at 6-15\,au, which is located interior to the orbit of the HR\,8799~e planet. The second is a much more spatially extended zone ($\approx$90 to $\approx$300\,au) of cooler dust ($T\approx 45$\,K) extending outward from the orbit  of the ``b'' planet. Hence, both debris components flank all four planets. Thus, with four giant planets with orbits residing in the cleared region between two dust belts, it has been suggested that the debris structure within the HR\,8799 system resembles a ``scaled-up'' version of our our own solar system \citep[e.g.][]{suetal2009,hugetal2018}.  

Importantly, careful characterization of the HR\,8799 debris structure, combined with dynamical arguments, have placed constraints on the masses of the individual planets. \citet{wiletal2018} used 1.3\,mm data from the \textit{Submillimeter Array} along with archival \textit{ALMA} data  \citep{booetal2016} to constrain the location of the inner edge of the outer debris belt to 104$^{+8}_{-12}$\,au. This much higher precision on the location of the belt edge, along with the sensitive mass dependence of HR\,8799~b on the spatial extent of the chaotic zone of the dust population, placed a strong constraint on the mass (5.8$^{+7.9}_{-3.1}$\,$M_{\rm Jup}$) of HR\,8799~b. This mass is consistent with the measurements of \citet{maretal2008}, who instead used evolutionary models of the luminosity of HR\,8799~b.

\begin{table*}
	\centering
	\caption{The exact $8$:$4$:$2$:$1$ mean motion resonance solution adopted here. This solution, from Go{\'z}dziewski \& Migaszewski (2020), is in astrocentric coordinates. The solution assumes initial co-planarity and a stellar (main-sequence) mass of $1.52M_{\odot}$.}
	\label{tab:param}
	\begin{tabular}{ccccc} 
		\hline
		 Parameter                          & HR 8799~e & HR 8799~d & HR 8799~c & HR 8799~b  \\
		\hline
		  Mass $(M_{\rm Jup})$              & 7.34688506 & 8.97059370 & 7.78986828 & 5.85290522\\ 
		  Semimajor axis (au)               & 16.21068245 & 26.59727940 & 41.27484337 & 71.42244964\\ 
		  Eccentricity                      & 0.14421803 & 0.11377309 & 0.05273512 & 0.01587597\\ 
		  Inclination (deg)                 & 26.55235715 & 26.55235715 & 26.55235715 & 26.55235715\\ 
		  Longitude of ascending node (deg) & 62.02658660 & 62.02658660 & 62.02658660 & 62.02658660\\ 		  		 
		  Argument of pericentre (deg)      & 111.05649395 & 28.35795849 & 92.75709369 & 41.35621703\\
		  Mean anomaly (deg)                & -23.88996589 & 60.75229572 & 144.93313134 & -47.73005711\\   
		\hline
	\end{tabular}
\end{table*}

\subsection{Dynamical evolution of HR 8799}

The architectures revealed by the observations have dynamical significance because of the very high masses of all four planets ($\approx 5-10 M_{\rm Jup}$), the wide planet-star separations of all four planets ($\approx 15-70$ au) and the relatively close planet-planet separations with respect to their stability boundaries. Initially, questions arose about how these planets were formed, particularly through fragmentation as opposed to core accretion \citep{dodetal2009,kraetal2010,merbat2010}. Further, even before the discovery of HR 8799~e, mean motion resonances between the planets were suspected as the driving forces behind the stability of the system \citep{gozmig2009,reietal2009,fabmur2010,marshalletal2010}. 

The discovery of HR 8799~e \citep{maretal2010} has prompted additional theoretical explorations of the formation and evolution of the system. Because disc fragmentation through gravitational instability is the favoured formation mechanism, this system has directly motivated several studies about the details of the process \citep[e.g.][]{baretal2011,boss2011,vorelb2018}. The apparent fragility of the system has also been investigated in the context of stellar birth clusters; HR 8799 might have been lucky to survive the cluster phase with its four planets intact  \citep{lietal2020}. Post-cluster evolution investigations not only considered the interplay between resonant behaviour and stability amongst the four planets \citep{gozmig2014,gozmig2018,gotetal2016,morkra2016}, but also the interactions between the planets and the debris belts in the system {\rev \citep{mooqui2013,gozmig2014,gozmig2018,gotetal2016,morkra2016}}.
   
However, the fate of the system beyond the main-sequence phase has not yet been investigated in detail. One potential reason is because the system is a rare example of a known exosystem which is not guaranteed to remain stable until the end of the main sequence. In fact, identifying a long-term stable solution given the observational constraints has been challenging. In a series of papers, \cite{gozmig2009,gozmig2014,gozmig2018,gozmig2020} have been updating and refining such a solution. Only in the most recent iteration \citep{gozmig2020} have the authors obtained an exactly periodic configuration which {\it guarantees} stability in the absence of external perturbations (e.g. from the Galactic environment) or internal perturbations (e.g. from stellar evolution). 

This solution allows us here to explore the post-main-sequence evolution of the system, albeit under a limiting ``most stable" assumption. The primary motivation for this study is simply to consider this unexplored aspect of the fate of this benchmark system. We do, however, also have a secondary motivation, one which is relevant to the mounting discoveries of planetary systems orbiting white dwarfs.

\subsection{An exemplar progenitor white dwarf planetary system}

Over 1000 white dwarf planetary systems are known through planetary debris that is detected in the photospheres of the stars \citep{dufetal2007,kleetal2013,kepetal2015,kepetal2016,couetal2019}. Given the frequency of this debris, accretion rates onto the white dwarfs, and planetary mass in the surrounding circumstellar environment \citep{koeetal2014,vanetal2015,farihi2016,rapetal2016,guretal2017,manetal2019,guietal2020,vanderbosch2020}, we assume that the debris arises primarily from the destruction of minor planets like asteroids as opposed to major planets like exo-Jupiters or exo-Earths. Nevertheless, we now know that such large planets can reach separations of under 0.1 au \citep{ganetal2019,vanetal2020}. Overall, major planets, minor planets, dust, gas and metallic debris have all been discovered around or in white dwarfs; for a recent review, see \cite{veras2021}.

The HR 8799 system is an exemplar progenitor white dwarf planetary system, for several reasons. First, the host star is an A-type main-sequence star, which represents the predominant progenitor stellar type for the known population of white dwarf planetary systems \citep{treetal2016,cumetal2018,elbetal2018,mccetal2020,barcha2021}. Further, the system contains multiple giant planets which will escape the reach of the host star's giant branch transformation \citep{musvil2012,adablo2013,norspi2013,viletal2014,madetal2016,ronetal2020}. These planets are also on the verge of instability, and instability during the white dwarf phase is necessary to perturb objects close to and later onto the star \citep{debsig2002,veretal2013a}. Finally, the HR 8799 system features multiple debris discs, and these could supply the metals that will be accreted by the white dwarf through interactions with the planets {\rev \citep{bonetal2011,debetal2012,frehan2014,musetal2018,smaetal2018,smaetal2021,veretal2021}}.

\subsection{Plan of paper}

In this paper, we model the full-lifetime post-cluster evolution of the four planets in the HR 8799 system by assuming that they currently reside in the stable, exact $8$:$4$:$2$:$1$ mean motion resonance orbital solution given by \cite{gozmig2020}. We include both internal and external forces which could destabilize the system in the form of stellar evolution and Galactic forces. In Section 2 we describe the orbital solution of \cite{gozmig2020}. Basic stability considerations are discussed in Section 3. Our numerical simulation setup and results are then presented in, respectively, Sections 4 and 5. We then conclude with a discussion (Section 6) and summary (Section 7).

\section{An exact resonance along the main-sequence}

We adopt the best-fitting orbital solution of \cite{gozmig2020} -- which ensures stability of the four planets in the absence of extra forces -- and list the solution in Table \ref{tab:param}. This configuration places the four planets in resonance immediately, such that at least one resonant angle between any two pairs of planets librates. In fact, the highlighted resonance achieved is a generalized $8$:$4$:$2$:$1$ Laplace resonance, reflecting the approximate orbital period ratios of planets b:c:d:e. More technically, this resonance expresses that the angle $\lambda_{\rm e} - 2 \lambda_{\rm d} - \lambda_{\rm c} + 2 \lambda_{\rm b}$ librates, where $\lambda$ denotes mean longitude.

The resonance is ``exact" in the sense of the configuration being strictly periodic with canonical elements rather than the amplitude of the Laplace resonance angle equalling zero. The libration amplitude is actually approximately $4^{\circ}$. \cite{gozmig2020} achieved the exact periodicity by appealing to the field of periodic orbit families, and in particular the four-planet case \citep{hadmic1981}. More recent applications of periodic orbit families in exoplanetary science have focussed on two-planet systems in the guise of the three-body problem \citep{antlib2018a,antlib2018b,antlib2019,antlib2020,voyant2018,voymou2018,mornam2019}, and to the type of white dwarf planetary system which HR 8799 might become \citep{antver2016,antver2019}.

\cite{gozmig2020} derived this solution by assuming that the mass of the star HR 8799 equals $1.52 M_{\odot}$ \citep{konetal2016}. For our study, this assumption may be crucial because the main sequence mass helps determine the duration and extent of giant branch mass loss, which in turn affects the stability of the system. Further, varying this mass, even by a few hundredths of a solar mass, may break the resonant configuration even on the main sequence. In this respect, the main-sequence lifetime then might affect the evolution.

Consequently, we have adopted three different stellar masses, straddling the uncertainties ($1.52 \pm 0.15 M_{\odot}$ from \citealt*{konetal2016}), although for most of our simulations we adopt the fiducial mass of $1.52M_{\odot}$. We use the {\tt SSE} code to model the stellar evolution \citep{huretal2000}. From this code, for main-sequence stellar masses of $1.37, 1.52, 1.67M_{\odot}$ respectively, we obtain main sequence lifetimes of (3.36, 2.62, 1.98) Gyr, giant branch lifetimes of (0.391, 0.284, 0.222) Gyr, and total fractions of mass lost during the giant branches of (0.591, 0.619, 0.643).

The current age of HR 8799 is uncertain, but as previously mentioned has been estimated to be on the order of tens of Myr. Our simulations are not highly sensitive to the current main-sequence age because the exactly periodic configuration is indefinitely stable in the absence of extra forces. Hence, we simply initiate HR 8799 at the zero-age main-sequence phase of evolution as a conservative limit.

\begin{figure}
\centerline{\ \ \ \ \ \ \ {\Large Main-sequence evolution}}
\centerline{}
\includegraphics[width=9cm]{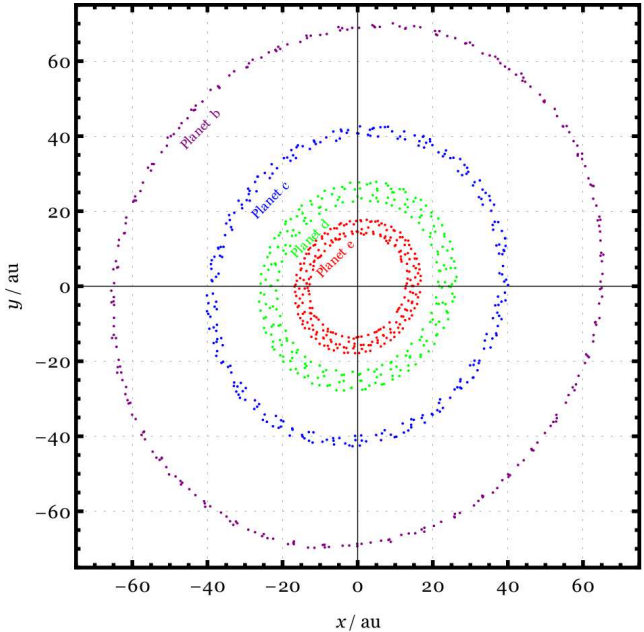}
\caption{
Astrocentric positions of the four HR 8799 planets in 10 Myr snapshots throughout the entire main sequence lifetime of the host star.
}
\label{Fig79}
\end{figure}

\begin{figure}
\includegraphics[width=9cm]{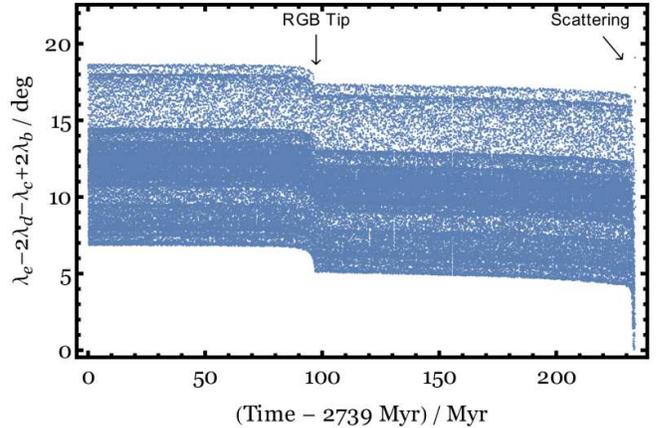}
\caption{
The evolution of the generalized Laplace resonant angle throughout the end of the giant branch phases, when the star is losing mass and the planets' orbits are expanding. The stellar mass loss at the tip of the red giant branch shifts the libration centre of the resonance by about $2^{\circ}$, after which the shift becomes more gradual. The subsequent evolution along the asymptotic giant branch eventually creates gravitational instability, breaking the resonance.
}
\label{Fig152}
\end{figure}

\begin{figure*}
\centerline{\ \ \ \ \ \ {\Large \underline{Representative case}}}
\centerline{}
\centerline{
\includegraphics[width=8.5cm]{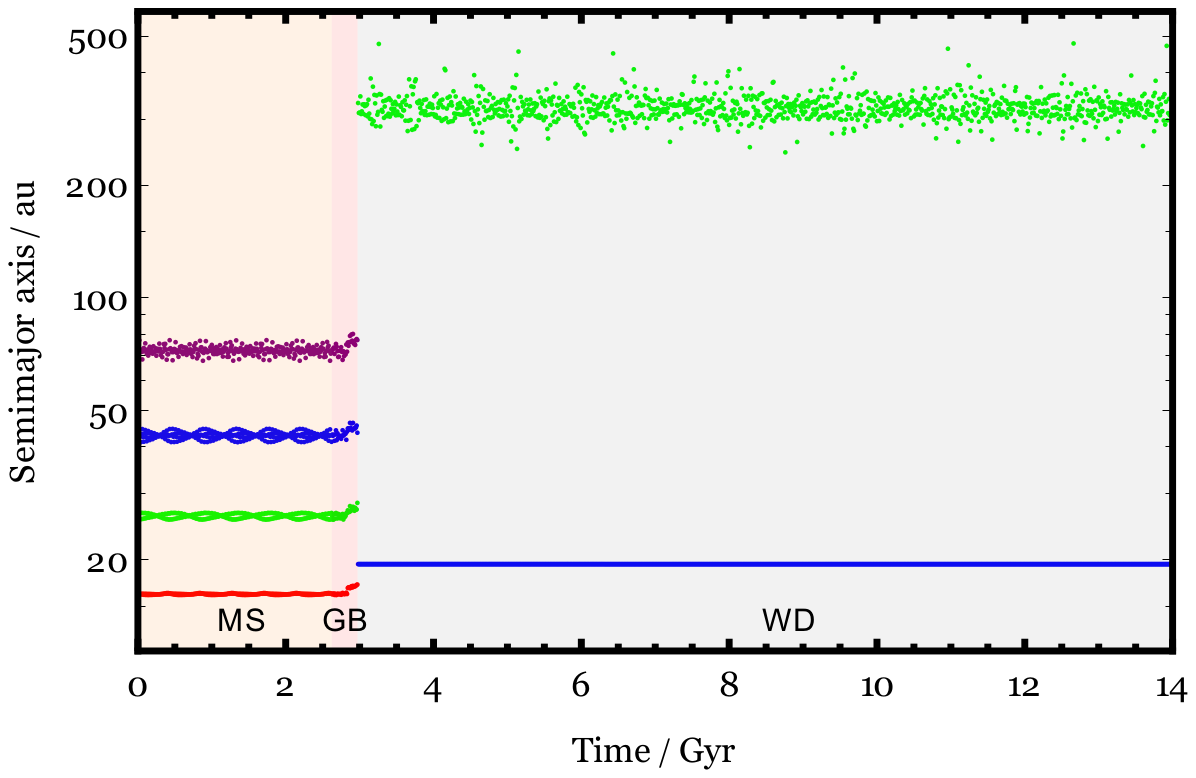}
\includegraphics[width=8.5cm]{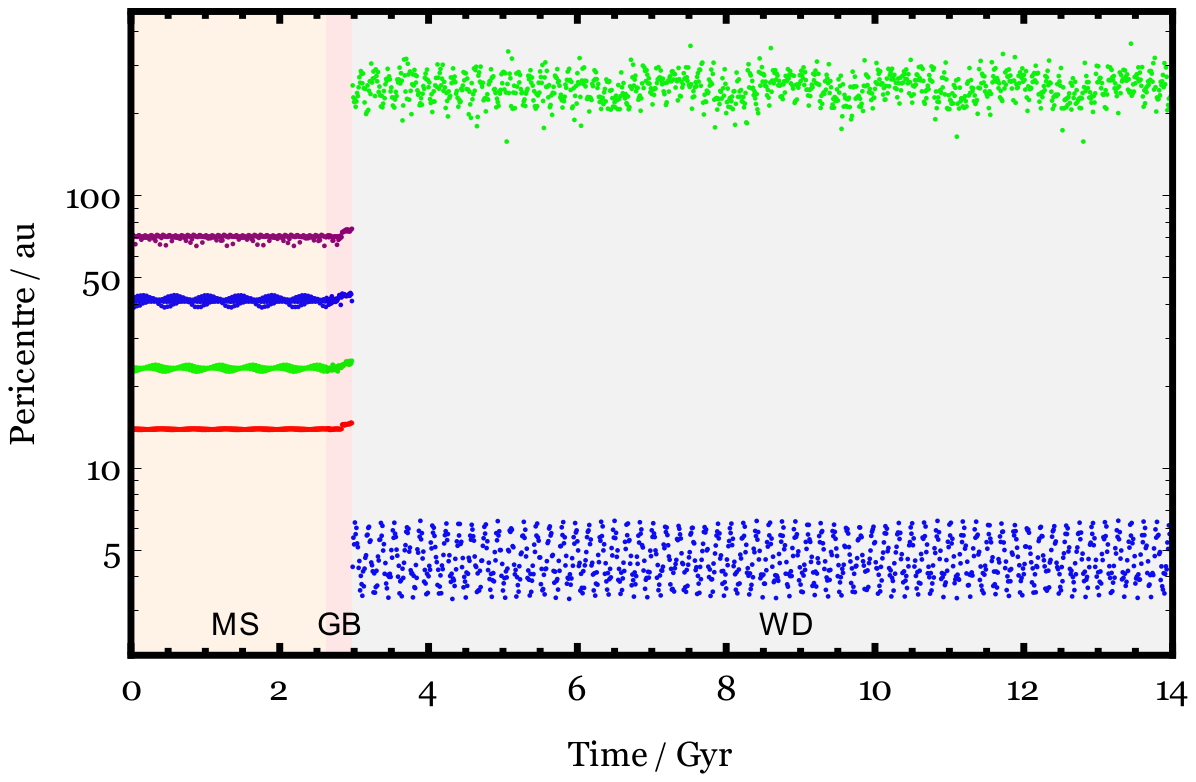}
}
\caption{
A representative evolution profile of semimajor axes and orbital pericentres, featuring the ejection of two planets (here planets b and e) just after the tip of the AGB, and the aftermath during the white dwarf phase. The abbreviations ``MS", ``GB" and ``WD" stand for main sequence, giant branch, and white dwarf respectively.
}
\label{Fig104}
\end{figure*}

\section{Basic stability considerations}

Having established the physical and orbital parameters that we will adopt for our simulations, now we discuss potential triggers for instability. One trigger is stellar flybys, which represent a potential danger throughout the lifetime of the system. The frequency and impact parameters of each flyby is unknown and probabilistic, but can be estimated through either analytic arguments or numerical simulations. 

For the HR~8799 system, the planet most susceptible to perturbations from flybys is HR~8799~b, the outermost planet. If this planet is sufficiently perturbed away from the exact periodic configuration, the result may be a (not necessarily immediate) scattering event. Another potential, but more unlikely, trigger for instability are Galactic tides. Nevertheless, we incorporate both stellar flybys and Galactic tides in our simulations during all phases of stellar evolution.

Those perturbations are external; the typically stronger perturbations are internal to the system, from the host star as it leaves the main sequence. The resulting increase in stellar radius and luminosity have just a negligible effect on the HR~8799 planets because of their wide orbits. However, the mass lost through stellar winds alters the gravitational potential of the system, expanding and stretching the planetary orbits \citep{omarov1962,hadjidemetriou1963}. 

The resulting variations in orbital eccentricity and argument of pericentre for a single planet within about $10^3$ au due to stellar mass loss can usually be considered negligible \citep{veretal2011}. However, the planet's semimajor axis would increase in proportion to the amount of mass lost: for our fiducial HR 8799 host star mass of $1.52M_{\odot}$, a mass loss fraction of 62 per cent would correspond to a semimajor axis increase factor of 2.6. If the mass loss is isotropic -- an assumption that we adopt -- then the inclination and longitude of ascending node of the orbit would remain fixed \citep{veretal2013b,doskal2016a,doskal2016b}.

In systems with more than one planet, stellar mass loss shifts stability boundaries such that a stable configuration on the main sequence becomes unstable on the giant branches or white dwarf phase \citep{debsig2002,veretal2013a,voyetal2013,musetal2014,musetal2018,veretal2018,maletal2020a,maletal2020b}. For systems of at least four planets (like HR 8799), instability in the entire system is easily triggered if any one pair of planets become unstable \citep{vergan2015,veretal2016,maletal2021}. Only if each planet is sufficiently separated from one another -- such as is the case with Jupiter, Saturn, Uranus and Neptune -- can the system remain stable throughout the giant branch and white dwarf phases \citep{dunlis1998,veras2016a,veras2016b,veretal2020,zinetal2020}.

Although the four HR 8799 planets are spread out more than Jupiter, Saturn, Uranus and Neptune, the significant difference in masses between the two sets of planets suggests that the HR 8799 planets might actually be more tightly ``packed" \citep[e.g.][]{rayetal2009}. Hence, whether all four of these planets can survive giant branch evolution is not immediately clear, particularly if they are stabilized through a resonance. Now we perform numerical simulations to explore this evolution.

\begin{figure*}
\centerline{\ \ \ \ \ \ {\Large \underline{Three-survivor case}}}
\centerline{}
\centerline{
\includegraphics[width=8.5cm]{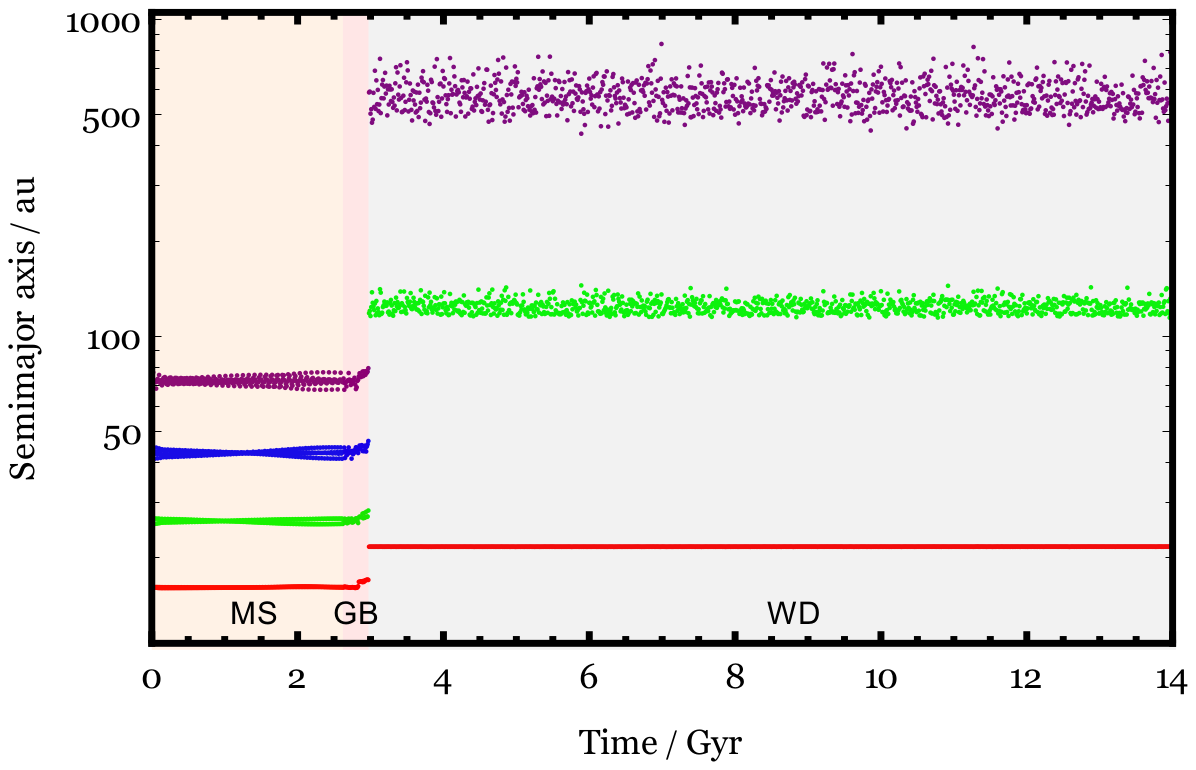}
\includegraphics[width=8.5cm]{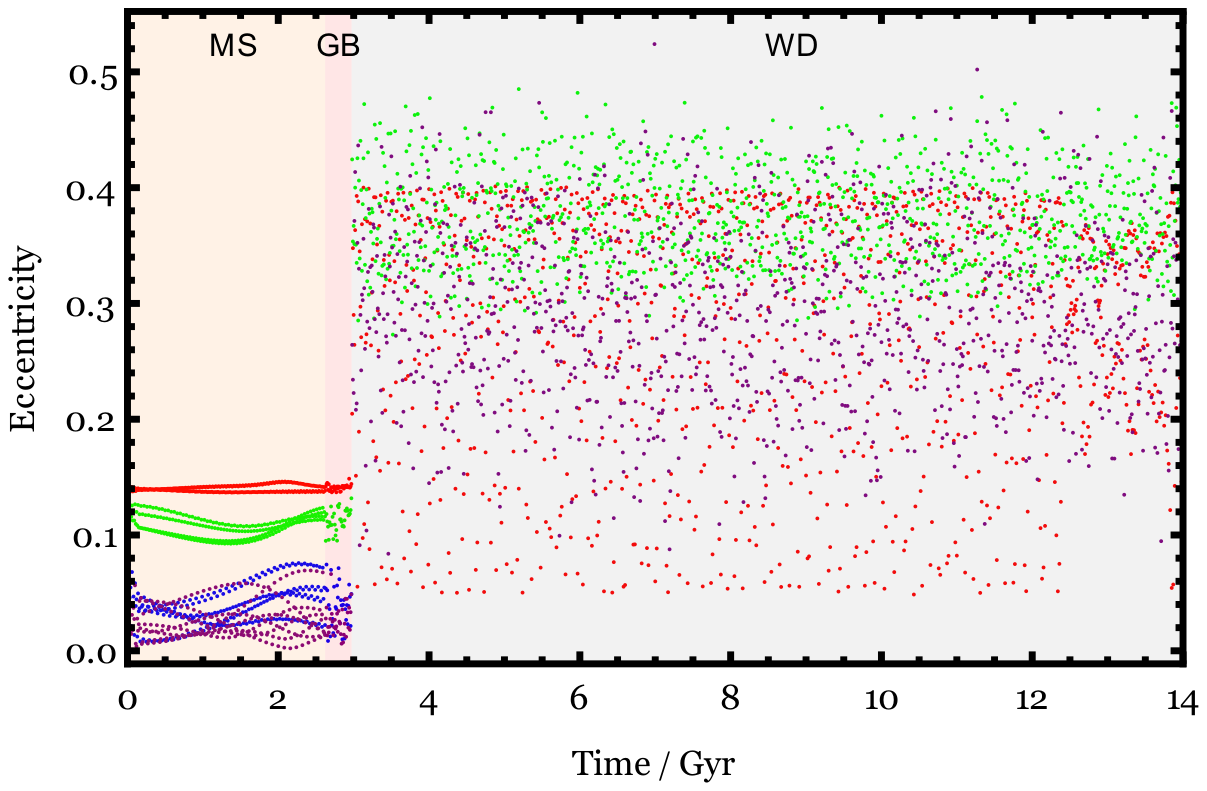}
}
\centerline{
\includegraphics[width=8.5cm]{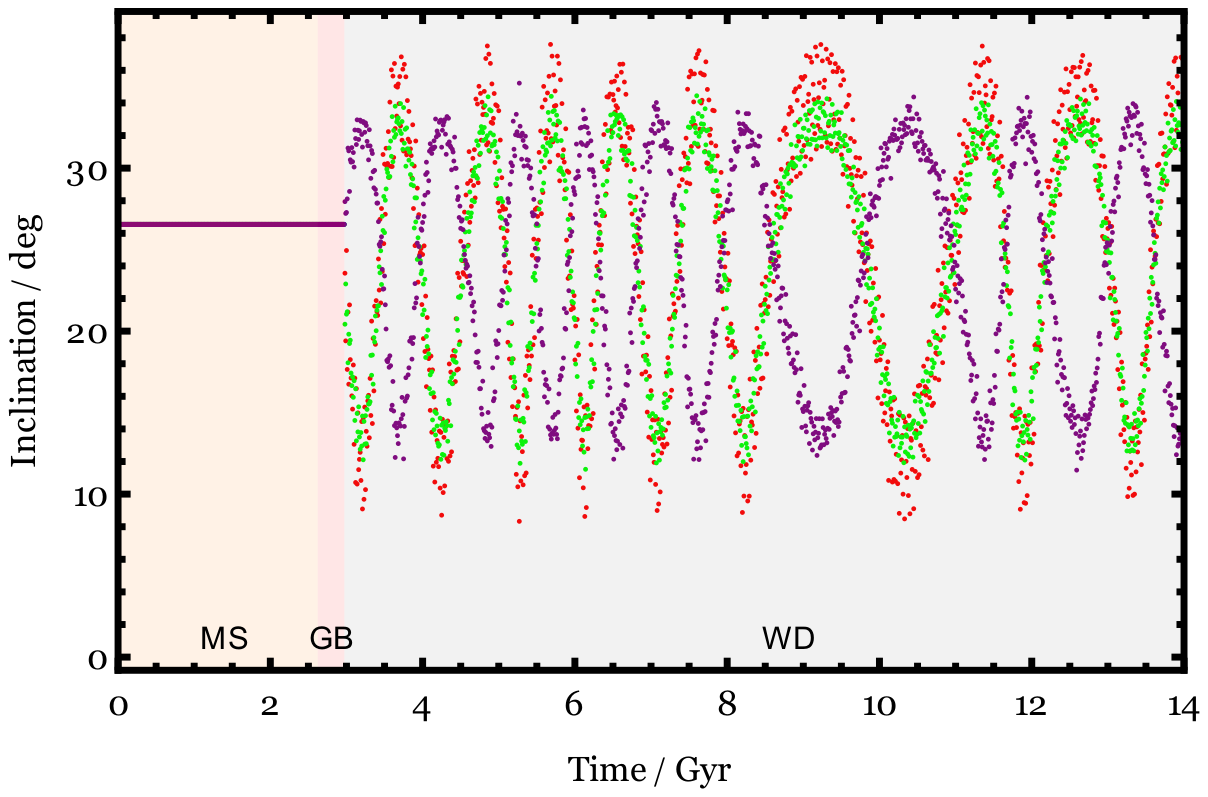}
\includegraphics[width=8.5cm]{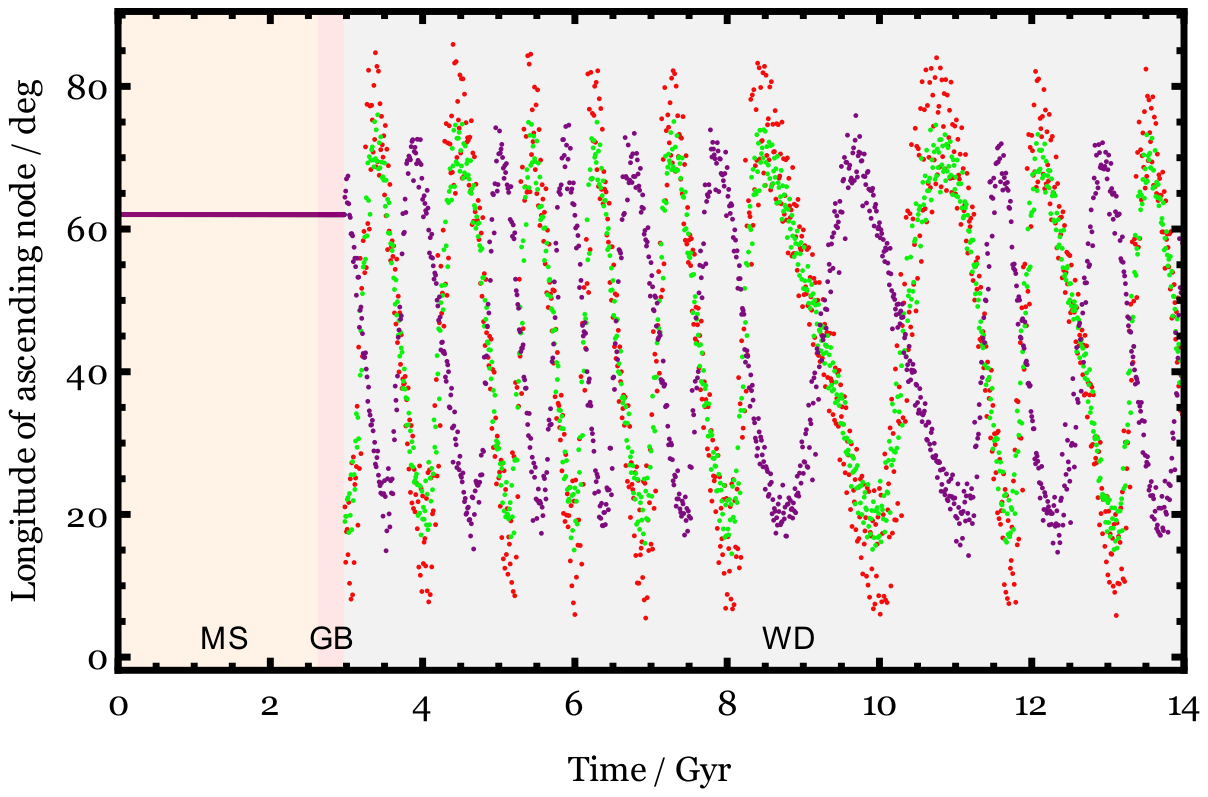}
}
\caption{
A rare case where three planets survive. Although all planets initially reside on coplanar orbits, stellar flybys create slight non-coplanarities. These are then exacerbated due to angular momentum exchange during the scattering event.
}
\label{Fig88}
\end{figure*}

\begin{figure*}
\centerline{\ \ \ \ \ \ {\Large \underline{Role of Galactic tides}}}
\centerline{}
\centerline{
\includegraphics[width=8.5cm]{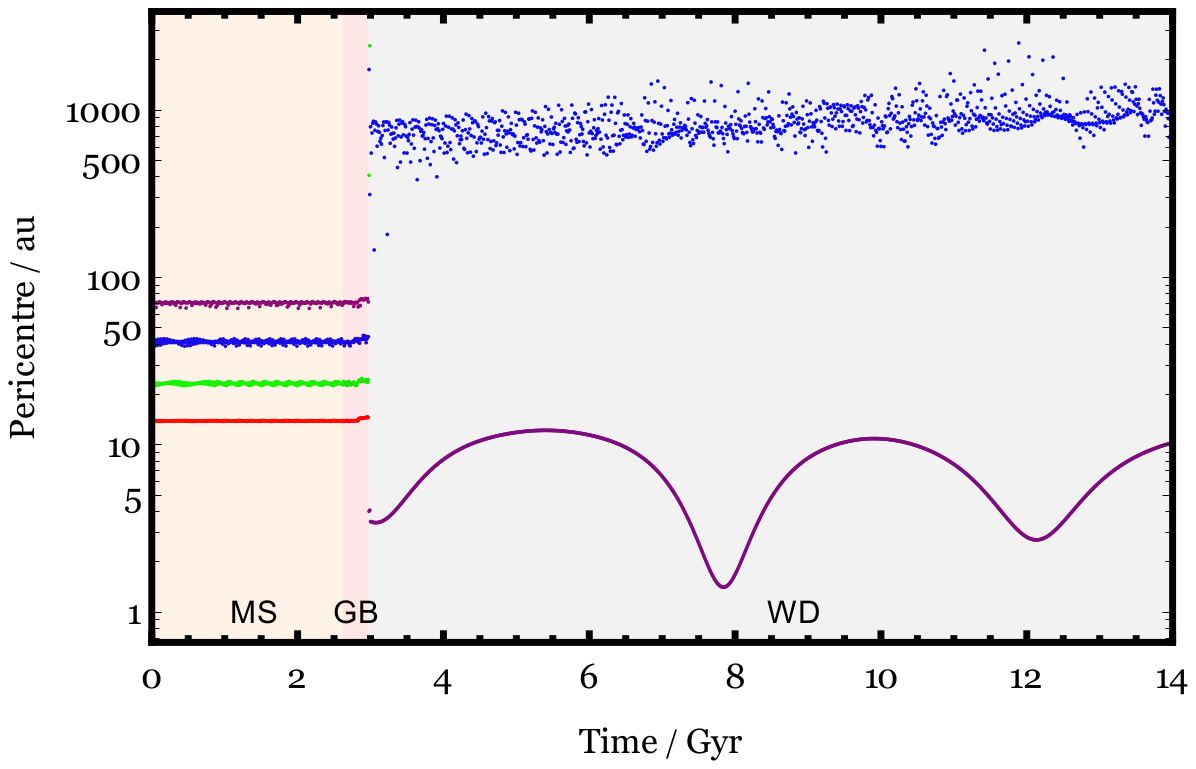}
\includegraphics[width=8.5cm]{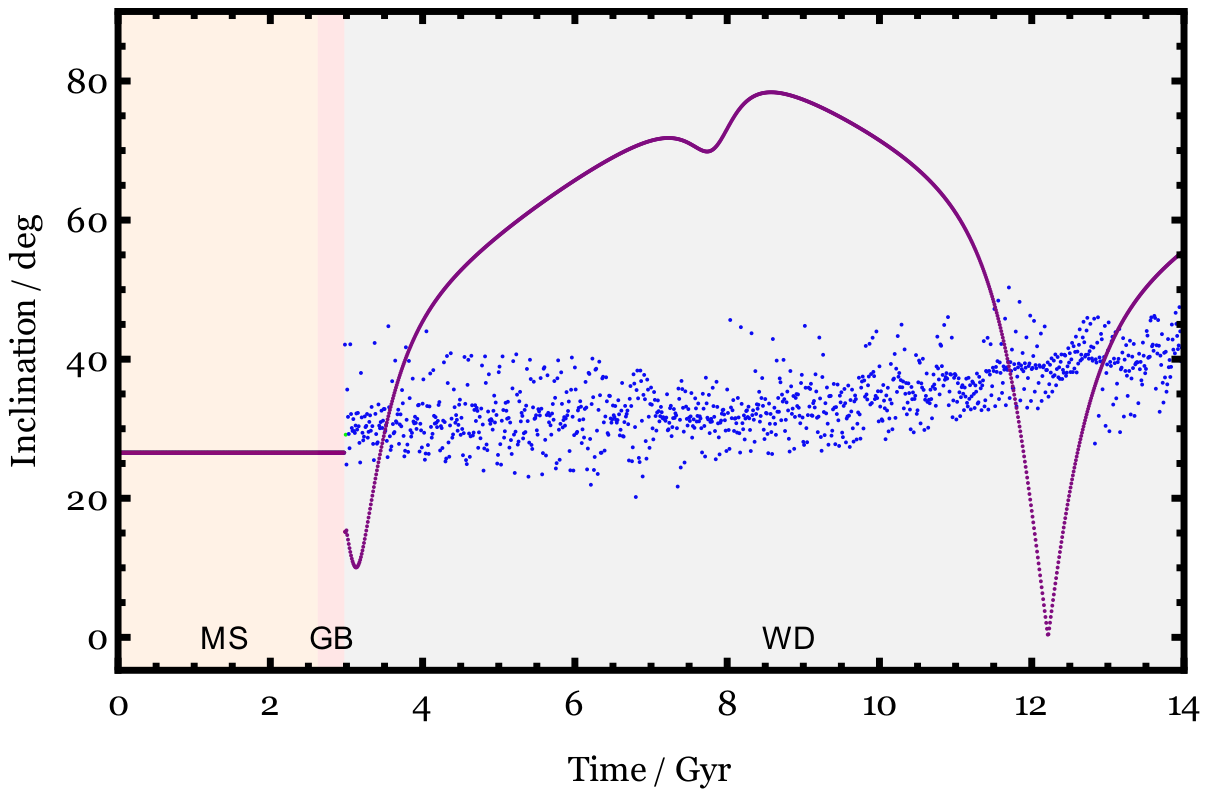}
}
\caption{
Evolution profiles which are affected by both Galactic tides (planet c, blue points) and von Zeipel-Lidov-Kozai effects (planet b, purple points) due to the large semimajor axis of planet c along the white dwarf phase and the oscillating values of inclination and eccentricity for planet b.
}
\label{Fig86}
\end{figure*}

\section{Numerical simulation setup}

In order to perform our numerical simulations, we use the RADAU integrator within the $N$-body code developed in \cite{veretal2013a} and subsequently modified in \cite{musetal2018}. This code is originally based on the {\tt Mercury} integration suite \citep{chambers1999}, but with the crucial difference that stellar evolution variations (with profiles from {\tt SSE}, \citealt* {huretal2000}, as described earlier) are incorporated into the integrations. The code also includes effects from stellar flybys and Galactic tides.    

The prescription we used for Galatctic tides is taken from \cite{vereva2013}, and assumes that the HR 8799 planetary system resides at an approximate distance of 8~kpc from the Galactic centre. For stellar flybys, we adopted the prescription from \cite{veretal2014a} but with an updated stellar mass function, stellar density and field encounter distribution from \cite{maretal2017}. The random introduction of stellar flybys provides a stochastic component to each simulation. Hence, although the initial masses and orbital parameters of each planet are equivalent across all simulations, the introduction of flybys creates a very slight perturbation which can change the fate of the system. 


When considering full lifetime simulations of planetary systems, one must balance accuracy and speed. In order to keep the HR~8799 planets locked in resonance throughout the main sequence, the integrator accuracy must be sufficiently high. Hence, we sampled four different accuracy tolerance values of $10^{-10}, 10^{-11}, 10^{-12}, 10^{-13}$ both to tease out any potential dependence, and to create another initial differentiating factor amongst different simulations. For each of these values and an initial mass of HR 8799 of $1.52 M_{\odot}$, we ran ten simulations for a total duration of 14 Gyr. We also ran ten simulations for each of the other two adopted initial stellar masses ($1.37 M_{\odot}$ and $1.67 M_{\odot}$) assuming an accuracy tolerance of $10^{-12}$. Finally, in order to track resonant behaviour due to mass loss, we ran one set of ten simulations with a very high output frequency (2340 yr) between 2.739 Gyr and 2.973 Gyr for an initial stellar mass of $1.52 M_{\odot}$ and an accuracy tolerance of $10^{-10}$.

In all simulations, we set the effective white dwarf physical radius to be $1 R_{\odot}$. This value, while hundreds of times larger than the star's actual physical radius, is a representative value of the white dwarf's Roche sphere. Any planet entering this Roche sphere will be destroyed. For the gas giants of the HR 8799 system, this value is conservative by a factor of a few \citep{verful2020}, but is large enough to prevent prohibitive timestep issues with the integration.

\begin{figure}
\centerline{\ \ \ \ \ \ \ \ \ \ {\Large \underline{Role of stellar flybys}}}
\centerline{}
\includegraphics[width=9cm]{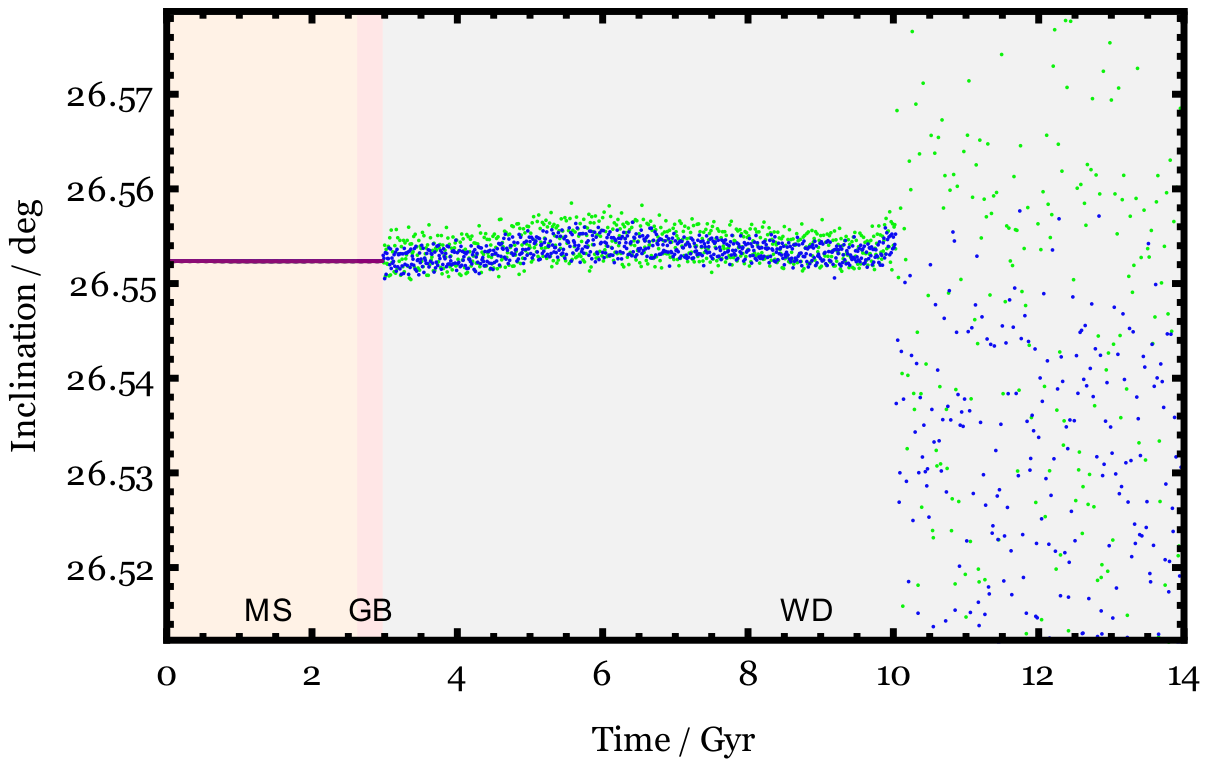}
\caption{
A visually demonstrable effect of a stellar flyby (at about 10 Gyr) on the change in inclination of the two surviving planets. 
}
\label{Fig87}
\end{figure}

\section{Numerical simulation results}

We break down our results into both notable individual system cases and ensemble system statistics. First, we highlight the result that in no case modelled did the resonant configuration break before the giant branch phases. Although Galactic tides and stellar flybys always altered the planets' orbital parameters, these variations were at a small enough level to maintain the resonance. Even changing the initial stellar mass from $1.52M_{\odot}$ to $1.37M_{\odot}$ or $1.67M_{\odot}$ did not break the resonance along the main sequence. Although altering the stellar mass rescales all of the planet-to-star mass ratios by the same factor, maintenance of resonance lock is not necessarily guaranteed; for HR 8799, however, resonance lock appears to be a sufficiently weak function of the planet-to-star mass ratio.

\subsection{Individual results}

We first show some representative evolutions, before considering particularly interesting cases. Figure \ref{Fig79} illustrates the astrocentric geometry of the planetary orbits throughout the main sequence, at 10 Myr snapshots. The exactly periodic configuration assumes co-planarity, which is only slightly broken by perturbations from stellar flybys. This slight shift is important because that can instigate a larger orbital plane divergence after a scattering event. Otherwise, because all of the orbital angular momentum for an exactly coplanar configuration would be normal to the orbital plane, in this case the planets' individual orbits would remain coplanar, and the incidence of planet-planet collisions would be artificially high.

Next, in Fig. \ref{Fig152}, we show a representative case of the evolution of the resonant angle $\lambda_{\rm e} - 2 \lambda_{\rm d} - \lambda_{\rm c} + 2 \lambda_{\rm b}$ as stellar mass is lost during the giant branch phases, for our fiducial initial stellar mass value ($1.52 M_{\odot}$). The time resolution of the plotted data points is 2340 yr, a small enough value to accurately assess resonant behaviour with respect to the giant branch timescale. The timespan of the plot covers the end of the red giant branch (RGB) phase, as well as the entire asymptotic giant branch (AGB) phase. The slight bump in the middle of the plot corresponds to the tip of the RGB, after which about 4 per cent of the stellar mass has been lost. The resonance breaks towards the right side of the plot, when instability sets in; this instability coincides with the significant and rapid mass loss that is characteristic of superwinds at the AGB tip \citep{vaswoo1993}. 

\begin{figure}
\centerline{\ \ \ \ \ \ \ {\Large \underline{Strongly nonuniform evolution}}}
\centerline{}
\includegraphics[width=9cm]{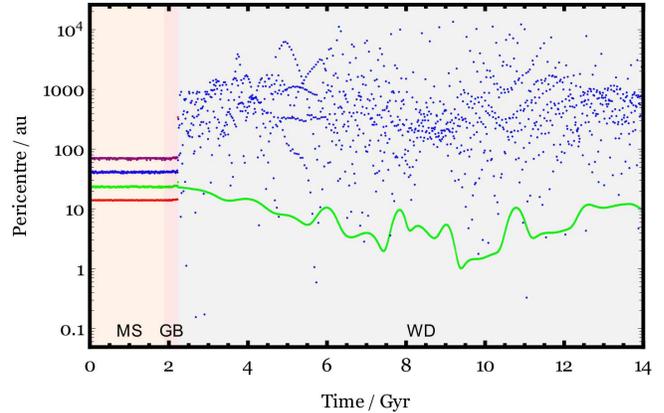}
\caption{
An example with highly nonuniform radial incursions of both planets c and d throughout white dwarf cooling. Such peregrinations are primarily the result of planet c's wide orbit, which stretches rapidly from Galactic tides. The evolution of both planets can access extant reservoirs of minor planets at different locations and times during the white dwarf phase.  
}
\label{Fig139}
\end{figure}

The most common outcome of the instability and resonant breaking is the ejection of two of the planets. The final planetary orbits of the survivors differ significantly from the final orbits which would be expected from mass loss-induced orbital expansion alone. 

Now we show some examples. In the following figures, we use osculating Jacobi coordinates to represent the orbital elements of the planets. We also highlight the orbital pericentre, which indicates the inner reach of a planet and hence its ability to and timescale for perturbing minor planets and debris into the eventual white dwarf. 

\begin{figure*}
\centerline{
\includegraphics[width=8.5cm]{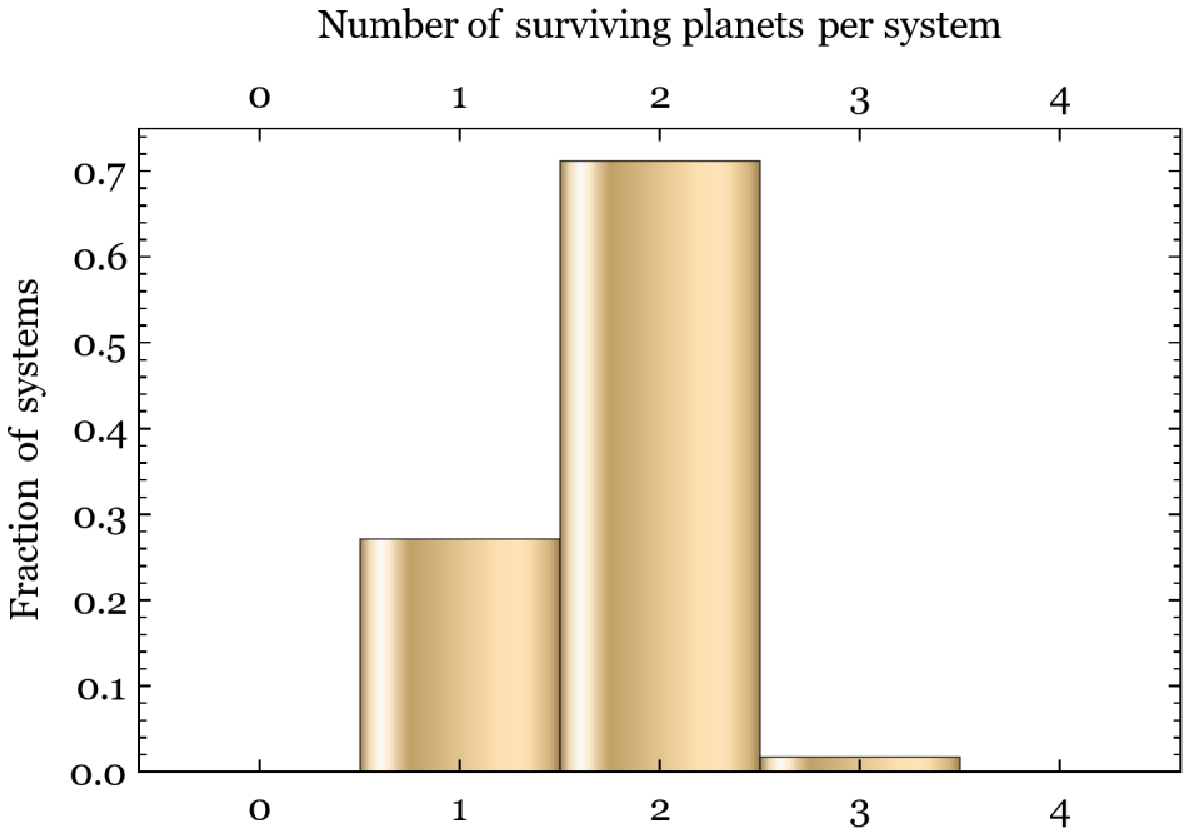}
\includegraphics[width=8.5cm]{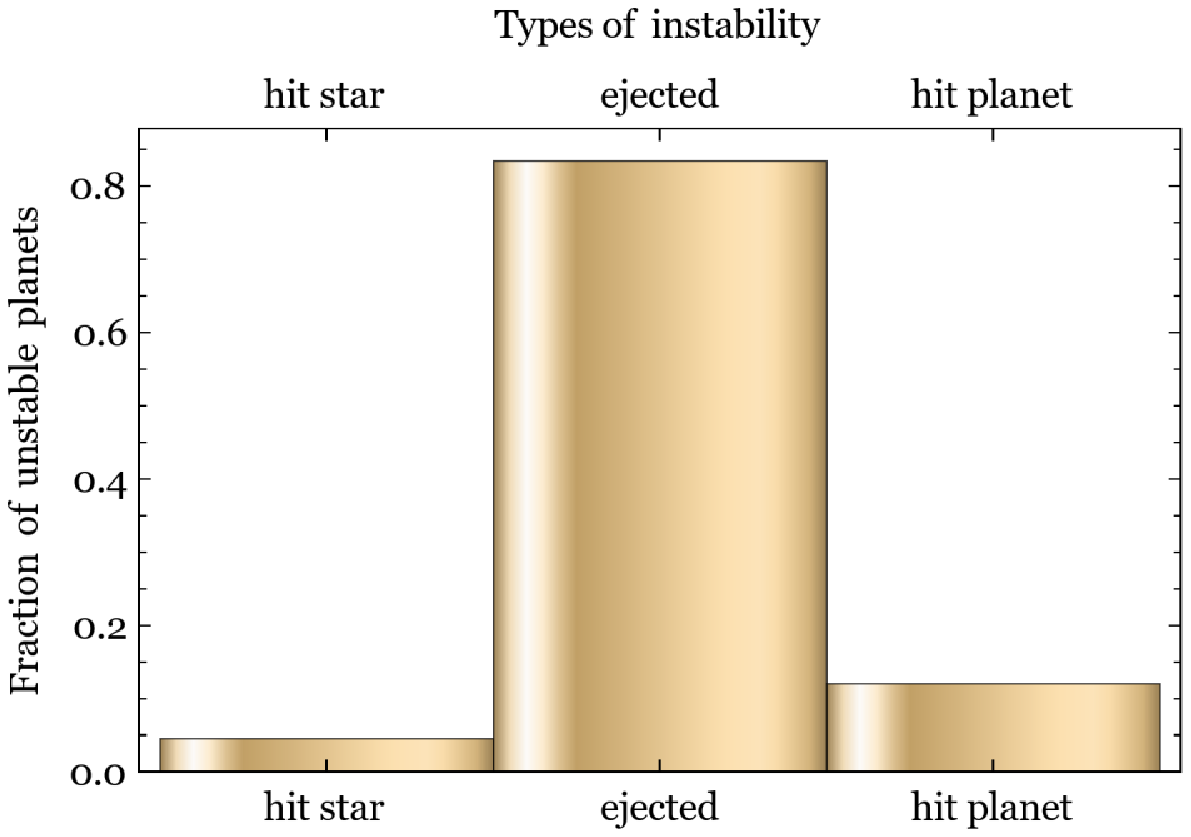}
}
\caption{
The number of planets which survive stellar evolution and the relative fractions of different instability types. The most common outcome is two planets being ejected.
}
\label{Fighist}
\end{figure*}

\subsubsection{Representative case}

One representative orbital evolution is shown in Fig. \ref{Fig104}. The semimajor axes of all four planets increase during the giant branch phases before an instability occurs at the tip of the AGB. The result of the instability is that planets b and e are ejected, and planets c and d swap orderings and are perturbed inward and outward. The semimajor axis of planet d becomes hundreds of au, making it more susceptible to Galactic tides and stellar flybys. The orbital pericentre of planet c becomes just 3-7~au, which is over 5~au smaller than the pericentre of planet e along the main sequence. The orbital evolution of both planets c and d remains relatively uniform throughout the white dwarf phase.

\begin{figure}
\includegraphics[width=9cm]{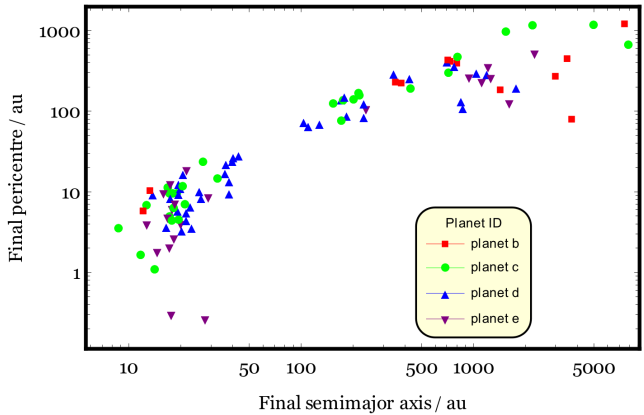}
\caption{
The semimajor axis and orbital pericentre of all planets which survived for 14 Gyr. The most massive planet, planet d, is the most robust survivor.  
}
\label{Figaq}
\end{figure}

\subsubsection{Three-survivor case}

Next we show a rarer outcome, one where three planets survive giant branch evolution (Fig \ref{Fig88}). In this case, a stellar flyby at 144 Myr (early on the main sequence) created a perturbation which was not strong enough to break the resonance, but strong enough to imprint a dynamical signature that manifests itself at the tip of the AGB. This signature allows planet c to be ejected, and all of the other planets to survive and maintain their ordering. Further, this flyby is strong enough to break the coplanarity of the orbits on the main sequence by a sufficient amount (although not noticeable on the plot), such that the instability triggers large (tens of degrees) oscillations in inclination and longitude of ascending node of the survivors. 

\begin{figure}
\includegraphics[width=9cm]{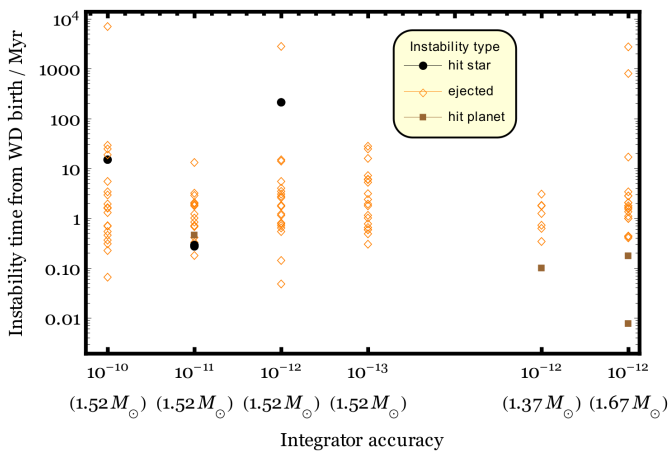}
\caption{
Instability times as a function of integrator accuracy and initial stellar mass. Nearly all instabilities are ejections and occur within a white dwarf cooling age of about 40 Myr.
}
\label{Figscatter}
\end{figure}

\subsubsection{Role of Galactic tides}

If the surviving outermost planet in a system is sufficiently distant, then Galactic tides start to play a role. In Fig. \ref{Fig86}, planet c is perturbed outward and subsequently harbours a semimajor axis of about 2,000 au. This value is high enough \citep{bonver2015} to create a gradual pericentre drift. The surviving inner planet, planet b, experiences secular nonuniform oscillations in eccentricity, which are manifested in the orbital pericentre plot. Because the oscillations are nonuniform, the planet may dynamically access different reservoirs of remnant planetary debris and minor planets throughout white dwarf evolution. The reason for the high amplitude secular oscillations is the von Zeipel-Lidov-Kozai effect, which effectively exchanges angular momentum between the eccentricity and inclination \citep{hampor2016,petmun2017,steetal2017,steetal2018,steetal2020,munpet2020,ocoetal2021}.

\subsubsection{Role of stellar flybys}

The strength of a stellar flyby cannot be characterized by the minimum close encounter distance with the star alone; in fact, the smallest closest encounter distance in any of our simulations (76 au) had no noticeable effect on the dynamics of that system. The more important measure is the direction-dependent closest proximity to any of the planets. Figure \ref{Fig87} illustrates a clear consequence of a stellar flyby, at a time of about 10 Gyr, on two surviving planets with semimajor axes at about 40 and 500 au. The flyby (with a closest encounter distance of about 785 au) increases the range of inclinations by one order of magnitude, albeit just from hundredths to a tenth of a degree. The figure also illustrates how much the initial inclination values are changed upon reaching the white dwarf phase.

\subsubsection{Strongly nonuniform evolution}

Example evolutions for the alternate host star mass cases are similar to the figures presented above. In one notable $1.67 M_{\odot}$ case (Fig. \ref{Fig139}), planets c and d survive the scattering event, and planet c is perturbed into an orbit with a semimajor axis of about 8,000 au and an eccentricity near unity. This planet is severely affected by Galactic tides, and the resulting perturbations on planet d, with a semimajor axis of about 30 au, produce a highly nonuniform orbital pericentre profile. The nonuniform radial incursions of both planets could potentially access and disperse reservoirs of minor planets towards the white dwarf at any time, including very late ($\approx 10$ Gyr) times. Accretion rates for such old white dwarfs have been measured \citep{holetal2018}.

\subsection{Ensemble results}

The wide variety of outcomes for individual simulations can be represented through a few ensemble illustrations. One relatively common outcome is the ejection of two of the four planets (Fig. \ref{Fighist}). The instabilities predominately produce ejections; rarely will a planet intersect the white dwarf's Roche radius, break up and be engulfed. 

Due to angular momentum balance, the ejection event is likely to perturb a surviving planet much closer to the white dwarf than at any point during the main sequence or giant branch phases. Figure \ref{Figaq} illustrates the final orbital pericentre versus semimajor axis for all planets which survive for a 14 Gyr timespan. Although the semimajor axes of nearly all these planets exceed 10 au, their orbital pericentres extend inward to about 0.3 au. The most common survivor is planet d, which is also the most massive of the four planets. In two cases the initially outermost planet, planet b, is perturbed to within about 10 au of the host star and survives.  

We also determine if integrator accuracy has a demonstrable effect on instability times in Fig. \ref{Figscatter}. The plot does not show any apparent correlation. The vast majority of instabilities occur within a cooling age (or time since the white dwarf was born) of 10 Myr. The ejection timescale is defined specifically as when the planet leaves the Hill ellipsoid of HR 8799 with respect to the Galactic centre \citep{vereva2013,veretal2014a} and is dependent on the direction of escape. Reaching this boundary typically takes on the order of Myrs, meaning that the gravitational scattering event for points on the plot could have occurred at or just after the tip of the AGB phase. 

This plot also illustrates that the planets in HR 8799 do not typically experience late gravitational scattering events, i.e. for cooling ages longer than 1 Gyr. Nevertheless, as demonstrated in Fig. \ref{Fighist}, the nonuniform orbital changes represent a continuous source of dynamical activity despite the lack of late ejections. Another tentative trend in the plot is the vertical scatter in the data points as a function of progenitor stellar mass: as this mass increases, so does the violence during the giant branch phases. More rapid and greater mass loss for the $1.67M_{\odot}$ case produces stronger scattering events, allowing planets to be more easily perturbed beyond $10^3$ au. At these distances, Galactic tides and flybys are more likely to instigate instabilities.

\section{Discussion}

{\rev Because our investigation considered only one set of initial conditions for the planets, our results do not necessarily represent the future evolution of the system. Further, even with this set of initial conditions, there is a small but finite probability that a very close stellar flyby could disrupt the system along the main sequence.}

{\rev In fact,} the striking variety of fates of HR 8799 from a single set of initial orbital parameters emphasizes the significant role external forces will play in the system's future evolution. {\rev Galactic tides could lead to qualitative secular behavioural changes for planets with post-scattered semimajor axes $\gtrsim 10^3$ au depending on the mutual inclination between the orbits of those planets and the Galactic plane (see e.g. Fig. \ref{Fig86} to see planet c's gradually increasing pericentre). Stellar flybys, however, affect systems differently, randomly generating short bursts of strong interactions, but only sometimes. Unfortunately, prospects for computing the future stellar dynamical properties of HR 8799's immediate neighbourhood are limited because of} the uncertainties in the stellar dynamics of Gyr-scale Galactic evolution. 

Nevertheless, the uncertainties about the minor planets in the system and their eventual locations are arguably greater. These bodies, currently unseen except for the dust they produce, are the potential metal polluters of the white dwarf. They can secularly or resonantly interact with the planets during giant branch evolution \citep{donetal2010} and experience grind-down and radiation blowout during this violent phase \citep{bonwya2010,zotver2020}. In addition, these bodies will be subject to the radiative Yarkovsky effect, which can propel them either outward or inward by tens or hundreds of au \citep{veretal2015,veretal2019a}. Some of asteroids closest to the star may also spin up to the point of rotational fission from the YORP effect \citep{veretal2014b,versch2020}, producing debris which then becomes more easily propelled by the Yarkovsky effect.

Wherever the minor planets and debris ultimately end up residing, we have shown the surviving planets typically and continually sweep out large areas encompassing the entire system from $\sim 0.1$ au to $\sim 10^5$ au. Based on this result, HR 8799 represents a viable future metal-polluted white dwarf. {\rev The pollution is likely to occur soon after the star becomes a white dwarf because the planets are so massive. Terrestrial-mass planets would allow minor planets to meander for multiple Gyr \citep{frehan2014,musetal2018}, whereas giant-mass planets would scatter the vast majority of minor planets within 1 Gyr \citep{veretal2021}. We then might expect HR 8799 to be polluted at cooling ages under about 1 Gyr.}

{\rev Although HR 8799 appears to contain all of the ingredients for the eventual white dwarf to be polluted, planetary systems as massive as HR 8799 are rare. Only a few systems are known to contain multiple massive giant planets, such as PDS 70 \citep{kepetal2018,hafetal2019} and TYC 8998-760-1 \citep{bohetal2020a,bohetal2020b}. Hence, although HR 8799 would likely represent a standard future polluted white dwarf, its surviving set of planets would not be typical.}

In our simulations, only rarely did the planets themselves enter the white dwarf Roche radius, or reach close enough to it to be significantly influenced by star-planet tidal effects \citep{verful2019,veretal2019b,ocolai2020}. {\rev Amongst the known population of four-planet systems \citep{maletal2021}, the fraction with close encounters between the white dwarf and planet is higher most likely because on average, those systems harbour smaller masses. Both \cite{vergan2015} and \cite{veretal2016} demonstrated that in systems with at least four planets, the smaller the planet masses, the longer the planets will meander and the greater the chance that they will enter the white dwarf Roche sphere.}

{\rev Our analysis of the HR 8799 system relies on a series of assumptions which we now discuss in more detail. The first is that the debris discs are not massive enough to alter the future stability of the four planets. \cite{mooqui2013} investigated this situation in detail (but not with the \citealt*{gozmig2020} orbital fit) and found that a sufficiently massive exterior disc may act as either a stabilizing or de-stabilizing influence on the planets. For discs to be effectual either way, they need to currently be at least as massive as Neptune. However, the disc masses in HR 8799 system remain unknown. The disc masses previously could have been larger, especially during and just after the protoplanetary disc phase.}

{\rev Another one of our assumptions was that the parent star emits mass isotropically and does not experience a natal kick during the transition to a white dwarf. Anisotropic mass loss would change the equations of motion for the variable-mass two-body problem \citep{veretal2013b,doskal2016a,doskal2016b}, but not qualitatively change the primary trigger for the instability, which is a shifting of stability boundaries due to mass lost \citep{debsig2002,veretal2013a}. Further, Fig. \ref{Fig152} indicates that the instability is triggered at the end of the AGB phase, which is also when the greatest anisotropy in stellar wind flow would occur. Overall then, we would not usually expect anisotropy to alter the instability timescale signficantly, but rather (if present) to change the fate of the system stochastically at a similar level to how weak random stellar flybys do so.}

{\rev A third assumption is that no additional planets reside in the system.} Hence, we now consider the prospects that as-yet unseen planets in HR 8799 exist. Following the initial discovery of the first three planets, \citet{hinetal2011} used interferometric imaging to search for any massive inner companions residing within 10 au of the system.  Although no new companions were found, these observations placed some of the strongest limits on companions in the innermost regions of the system, including the region within about 6\,au that is relatively cleared of dust. 

However, as more advanced coronagraphs have been deployed \citep[e.g.][]{maietal2015}, and sophisticated image processing techniques have matured, more recent observations \citep[e.g][]{curetal2014,wahetal2021} have placed comparable limits on an inner fifth companion in the system. Similarly, the morphology of the outer circumstellar debris structure is perhaps the best indicator of where additional planets may reside at wide separations. Given that modern ground based observations easily achieve sensitivities in the HR\,8799 system down to a few Jupiter masses within about 200\,au, the masses of any additional companions in the system will likely be $\lesssim$1-2\,$M_{\rm Jup}$.  Such low masses will be easily attainable given the sensitivity of the upcoming \textit{James Webb Space Telescope} \citep{caretal2021}.

\section{Summary}

We have modelled the fate of the HR 8799 planetary system under the assumption that all four planets will survive until the end of the main sequence in a resonant configuration. Because of the stochastic nature of the system, markedly different outcomes are generated by just tweaking the integrator accuracy or introducing external effects; stellar flybys and Galactic tides can play a significant role in the evolution during the white dwarf phase after one of the planets has been scattered out to a distance of hundreds or thousands of au. Because HR 8799 is an A-type star, it represents the typical progenitor of the white dwarfs currently observed in the Galaxy. With the star's four planets, debris belts, and dynamical activity during the white dwarf phase, HR 8799 contains all of the necessary ingredients to become a debris-polluted white dwarf.

\section*{Acknowledgements}

{\rev We thank the reviewer for their insightful comments and perspective, which have improved the manuscript.}
DV gratefully acknowledges the support of the STFC via an Ernest Rutherford Fellowship (grant ST/P003850/1). 

\section*{Data Availability}

The simulation inputs and results discussed in this paper are available upon reasonable request to the corresponding author.

\label{lastpage}
\end{document}